\documentclass[aps,prl,twocolumn,nofootinbib,superscriptaddress,floatfix]{revtex4-1}  
\usepackage{array,multirow,graphicx}  

\usepackage{amsmath}
\usepackage{amssymb}   
\usepackage{color}
\usepackage{hyperref}
\usepackage{ulem}
\usepackage{natbib}
\usepackage{enumitem}

\newcommand{\ltsima} {$\; \buildrel < \over \sim \;$}
\newcommand{\gtsima} {$\; \buildrel > \over \sim \;$}
\newcommand{\lta} {\lower.5ex\hbox{\ltsima}}
\newcommand{\gta} {\lower.5ex\hbox{\gtsima}}

\newcommand\be{\begin{equation}}
\newcommand\ee{\end{equation}}

\begin{document}

\title{k-cut Cosmic Shear: Tunable Power Spectrum Sensitivity to Test Gravity}

\author{Peter L. Taylor}\email{peterllewelyntaylor@gmail.com}
\affiliation{Mullard Space Science Laboratory, University College London, Holmbury St.~Mary, Dorking, Surrey RH5 6NT, UK}
\author{Francis Bernardeau}
\affiliation{UPMC - CNRS, UMR7095, Institut d’Astrophysique de Paris, F-75014, Paris, France}
\affiliation{CEA - CNRS, URA 2306, Institut de Physique Th$\acute{e}$orique, F-91191 Gif-sur-Yvette, France}
\author{Thomas D.~Kitching}
\affiliation{Mullard Space Science Laboratory, University College London, Holmbury St.~Mary, Dorking, Surrey RH5 6NT, UK}
	
\date{\today}

\begin{abstract}
If left unchecked modeling uncertainties at small scales, due to poorly understood baryonic physics and non-linear structure formation, will significantly bias Stage IV cosmic shear two-point statistic parameter constraints. While it is perhaps possible to run N-body or hydrodynamical simulations to determine the impact of these effects this approach is computationally expensive; especially to test a large number of theories of gravity. Instead we propose directly removing sensitivity to small-scale structure from the lensing spectrum, creating a statistic that is robust to these uncertainties. We do this by taking a redshift-dependent $\ell$-cut after applying the Bernardeau-Nishimichi-Taruya (BNT) nulling scheme. This reorganizes the information in the lensing spectrum to make the relationship between the angular scale, $\ell$, and the structure scale, $k$, much clearer compared to standard cosmic shear power spectra -- for which no direct relationship exists. We quantify the effectiveness of this method at removing sensitivity to small scales and compute the predicted Fisher error on the dark energy equation of state, $w_0$, for different $k$-cuts in the matter power spectrum.
\end{abstract}
%



\maketitle

\section{Introduction}
The cosmic shear signal is sensitive to the geometry and density field of the low redshift Universe, precisely where dark energy becomes important. This makes it an ideal probe of gravity on cosmic scales. Cosmic shear is a `clean' probe in the sense that it directly traces the density field without having to assume a biased tracer model, as in galaxy clustering studies~\cite{laureijs2011euclid}. Furthermore, cosmic shear extracts information about both the Newtonian potential $\Psi$ and the curvature potential $\Phi$~\cite{simpson2012cfhtlens}.
\par Nevertheless cosmic shear comes with its own set of unique theoretical challenges including the challenge of shape measurements~\cite{zuntz2017dark, massey2007shear,kitching2013image} and sensitivity to changes in the small scale behavior of the matter power spectrum. In this paper we propose a solution to the later of these issues, the so called {\it Small Scale Sensitivity Problem}, namely that the shear signal is sensitive to poorly understood small scale structure -- down to $k = 7 \text{ }h \text{ Mpc} ^{-1}$~\cite{taylor2018preparing}. Modeling the impact of baryons and nonlinear structure formation at these scales, to the level of accuracy required for Stage IV experiments \cite{albrecht2006report,laureijs2010euclid,spergel2015wide, anthony4836large}\footnote{\url{http://euclid-ec.org}} \footnote{\url{https://www.nasa.gov/wfirst}} \footnote{\url{https://www.lsst.org}}, presents a formidable challenge.
\par To attempt to overcome this problem a large amount of work has been devoted to the brute force $N$-body simulation approach, which is used in all current shear two-point statistic studies. In this paradigm, power spectrum emulators~\cite{halofit, lawrence2010coyote} or calibrated halo model codes~\cite{mead2015accurate} are trained on a large number of $N$-body simulations, that sample cosmological parameter space. Coupling the emulators to a lensing code~\cite{cosmosis, taylor2018testing, schneider2002analysis} to compute shear two-point statistics enables rapid Markov Chain Monte Carlo (MCMC) parameter inference as in~\cite{heymans2013cfhtlens,troxel2017dark,kohlinger2017kids}. Nevertheless, current state-of-the-art emulator codes are not sufficiently accurate for Stage IV lensing surveys~\cite{taylor2018preparing, eifler2015accounting, semboloni2011quantifying,mead2015accurate, Huang:2018wpy}.  
\par Although it may be possible to supplement the brute force approach by marginalizing out the small scale information, as proposed in~\cite{eifler2015accounting}, it is infeasible to run a large number of $N$-body simulations to test {\it all} theories of gravity, without using the untested assumption that nonlinear and baryonic feedback is cosmology and model independent. Even if this was possible, the standard approach is still far from ideal. Since cosmic shear is so sensitive to small scales ($\sim 50 \%$ of the information comes from scales below $k_{\text{cut}} = 1$ $h$ $\text{Mpc} ^ {-1}$ ~\cite{taylor2018preparing}), unknown or unmodeled baryonic physics at even smaller scales could easily bias the cosmological inference. 
\par We propose a cleaner geometric solution to the {\it Small Scale Sensitivity Problem} which efficiently cuts out the lensing spectrum's sensitivity to small scale structure, allowing for a tuneable $k$-mode sensitivity. We refer to this procedure as {\it $k$-cut cosmic shear} which has two parts. First, we apply the Bernardeau-Nishimichi-Taruya (BNT) nulling scheme~\cite{bernardeau2014cosmic} which reorganizes the information originally binned in the source plane to bins in the lens plane, then since each bin labels a {\it lens} redshift range, taking an angular scale cut also removes sensitivity to large-$k$ (small scales). See~\cite{simpson2015enhancing} for an alternative approach that reduces sensitivity to small scales.
\par In the next section we review the $BNT$ nulling scheme (we refer the reader to~\cite{bernardeau2014cosmic}, which provides the main theoretical backbone of this work, for more details). Then we introduce $k$-cut cosmic shear. In the remaining sections we discuss its effectiveness and future prospects. The main results are summarized in Figures~\ref{fig:f1}-\ref{fig:w0}.  

\section{k-cut Cosmic Shear}
Suppose we wish to remove from the projected lensing spectrum, $C_\ell$, contributions from structure smaller than some scale, denoted by a $k$-mode. If we lived in a `shell universe', where all the matter lay at a distance $r$, then the Limber relation~\cite{loverdelimber} tells us that we could simply cut angular scales $\ell > kr$. Unfortunately in the real Universe, the lensing kernel is broad, so lenses across a wide range of distances and scales contribute power to the same $\ell$-mode, which means such a strategy will not work by itself~\cite{taylor2018preparing}.  
\par We now review the steps of the Bernardeau-Nishimichi-Taruya (BNT)~\cite{bernardeau2014cosmic} formalism which re-weights the standard tomographic $C_\ell$ so that each bin contains information only about the {\it lenses} inside a small redshift range. It is then a simple extension to apply the Limber argument in each bin to cut sensitivity to small scales. 
\par To begin, suppose there are a discrete number of source planes at radial distances $r_i$. Then the weighted convergence, $\tilde \kappa$, can be written as :
\begin{equation}
\tilde \kappa = \frac{3 \Omega_m H_0 ^ 2}{2c ^2}  \int ^ {r_i} _0 \text{d}r   \frac{ \delta \left( r \right)}{a \left( r \right)} w \left( r \right)
\end{equation}
where $\delta \left(r \right)$ and $a \left(r \right)$ respectively give the local matter overdensity and scale factor of the infinitesimal lens at the radial distance $r$. Here:
\begin{equation}
w \left( r \right) = \sum_{i, r_i >r} p_i \frac{r_i - r}{r_i},
\end{equation}
where $\{ p_i\}$ are a set of weights~\cite{bernardeau2014cosmic}.
\par If we now assume there are just three discrete source planes: $r_1 < r_2 < r_3$, then the key step in the BNT nulling scheme is to construct constant weights $p_i$ so that $w \left( r \right) = 0$ for $r<r_1$. Clearly `lenses' at $r ' > r_3$ do not contribute to $\tilde \kappa$. Together these observations imply the weighted convergence is only sensitive to lenses that lie in the radial range $r \in \left[r_1,r_3 \right]$. 
\par This argument is generalized in Section~2.2 of~\cite{bernardeau2014cosmic} to an arbitrary number of source planes to construct a weighting matrix, $M$, which has the property that for each tomographic bin in the weighted lensing spectrum $\tilde C_\ell = MC_{\ell}M^T$\footnote{This applies to both the shear and convergence spectrum because the two full-sky spectra are related by $C_\ell ^{\kappa \kappa}  = \frac{\ell ^ 2 (\ell + 1) ^2 }{(\ell + 2)(\ell + 1)\ell(\ell - 1)} C_{\ell}^{\gamma \gamma}$.} is only sensitive to lensing structure in a small redshift range.  The shot noise spectrum, $N_\ell$, must also be consistently re-weighted and it is mapped to $ \tilde N_{\ell} = MN_{\ell}M^T$. Crucially the matrix, $M$, has ${\rm det}\left( M\right) = 1$, so the signal-to-noise remains unchanged.
\par Now in each re-weighted tomographic bin, $i$, if we choose the minimum lens distance, $r^{min}_i$, cutting all $\ell$-modes such that $\ell > k_{\rm cut} r^{min}_i$ will remove sensitivity to all scales smaller than $k_{\rm cut}$. 
\par In our analysis, we use the formalism of Section 2.2 in~\cite{bernardeau2014cosmic} to construct the BNT weight matrix, $M$, from $10$ tomographic bins each containing the same number of galaxies. The assumed radial distribution of galaxies $n(z)$ is given in the Appendix. Then for each BNT reweighed bin, $i$, rather than cutting $\ell > k_{\rm cut} r^{min}_i$, we instead use the mean distance to each weighted bin. This means that we do not have to cut the first bin entirely. Although using the mean rather than the minimum distance will not completely remove sensitivity to all $k$ above the target cut we show this has negligible impact. We must also assume a fiducial cosmology to go from the redshift, $z$, to co-moving distance $r (z)$. This is given in the Appendix. Finally in the cross-correlation between bins we take whichever $\ell$-cut is smaller.
\par We refer to the joint procedure of BNT annulling and applying a lens-redshift dependent angular scale cut as {\it $k$-cut cosmic shear}. 

\section{Fisher Matrix Formalism} \label{sec:fisher}
\par We now review the Fisher matrix formalism that we use to evaluate the sensitivity of the standard $C_\ell$ analysis, BNT cosmic shear, and $k$-cut cosmic shear to the matter power spectrum and compare constraints on the dark energy equation of state. 
\par For a set of parameters $\{ \theta_i \}$ the Fisher matrix for cosmic shear is given by:
\begin{equation} \label{eqn:fish}
F_{\alpha\beta} = \sum_{\ell} \frac{2\ell+1}{2} Tr \left[  C_{\ell}^{-1}  C_{{\ell},\alpha}  C_{\ell}^{-1}  C_{{\ell},\beta} \right],
\end{equation}
where $C_{{\ell},\alpha}$ denotes the derivative with respect to parameter $\theta_{\alpha}$. This $C_{\ell}$ includes both the signal and the noise contribution defined in equations~\ref{eq:c_l} and \ref{eq:Noise} in the Appendix. The lensing spectra are computed using {\tt GLaSS}~\cite{taylor2018testing}  which is integrated into the {\tt Cosmosis}~\cite{cosmosis} modular cosmology package. Details of the lensing spectra calculation are given in the Appendix.
\par To measure the sensitivity of cosmic shear to the matter power spectrum, we follow the analysis of~\cite{taylor2018preparing} which we now review. First we divide the matter power spectrum $P \left( k, z \right)$, into logarithmically and linearly spaced grid cells in $k$ and $z$, respectively. We then compute the fractional amplitude change in the power spectrum inside each grid cell $g$:

\begin{equation}
	P_g \left(k,z, \mathcal{A} \right)  \equiv
    \begin{cases}
      \left( 1 + \mathcal{A}\right) P \left(k,z \right) & \text{if $(k,z)$ in cell $g$  }\\
      P \left(k,z \right) & \text{otherwise},\\
    \end{cases}
\end{equation}
where $\mathcal{A}$ is a fixed small amplitude change.  The two sided derivative is:
\begin{equation}
 C_{\ell,g} = \frac{ C_\ell \left[ P_g \left(k,z, \mathcal{A} \right) \right] - C_\ell \left[ P_g \left(k,z, -\mathcal{A} \right] \right)}{2 \mathcal{A}},
\end{equation} 
where $,g$ denotes the derivative with respect to amplitude of cell $g$.
Putting this into equation~\ref{eqn:fish} gives the Fisher matrix $F$ for the matter power spectrum grid cells. Then the sensitivity to power spectrum cell $g$ is defined by the inverse error, $\sigma^ {-1} (\mathcal{A}_g)$, given by:
\begin{equation} \label{eqn:error}
\sigma^ {-1} (\mathcal{A}_g) = \frac{1}{\sqrt{(F ^ {-1})_{gg}}}.
\end{equation}
\par In a similar fashion we compute the error on $\Omega_m$, $\tau$, $\Omega_b$, $H_0$, $\sigma_8$ and $w_0$ in a flat universe. We then compare the relative change in marginalised constraints on the dark energy equation of state $w_0$ for a given analysis denoted by $\sigma (w_0)$ relative to the standard $C_\ell$ analysis where we denote the error as $\sigma_{\rm fid} (w_0)$. We do not compute the constraints on $w_a$ because we have found that this can be sensitive to exactly how the derivative is defined.  

\section{Results}
Using the formalism presented in the previous section we compute the sensitivity of different analyses to regions of the matter power spectrum and compare the constraints on the dark energy equation of state parameter $w_0$. In particular we consider: 
\begin{itemize}
\item{the standard cosmic shear $C_{\ell}$ approach with a large constant $\ell_{max}$;}
\item{a BNT re-weighed $C_{\ell}$ analysis with no $\ell$-cuts;}  
\item{$k$-cut cosmic shear for target $k_{\rm cuts}$ of the form $k_{\rm cut} = A_{cut}$ (redshift independent) and $k_{\rm cut} = A_{\rm cut} (1 + z) ^2$. In the former case we use $A_{\rm cut} \in [0.64, 1.94, 3.38]$ as representative. Meanwhile in the later case we consider $A_{\rm cut} \in [0.2, 0.6, 2]$ which roughly follows the redshift evolution of the highest $k$-mode in the linear regime, a $k$-value in the quasi-linear regime and a $k$-value in the fully non-linear regime.}
\end{itemize}
\par
\begin{figure*}
	\begin{minipage}{1.0\textwidth}
		{\bf Standard}\;\;\:\:\:\;\qquad\qquad\qquad\qquad\qquad{\bf BNT weighted}\\ 
        \includegraphics[width=5.5cm]{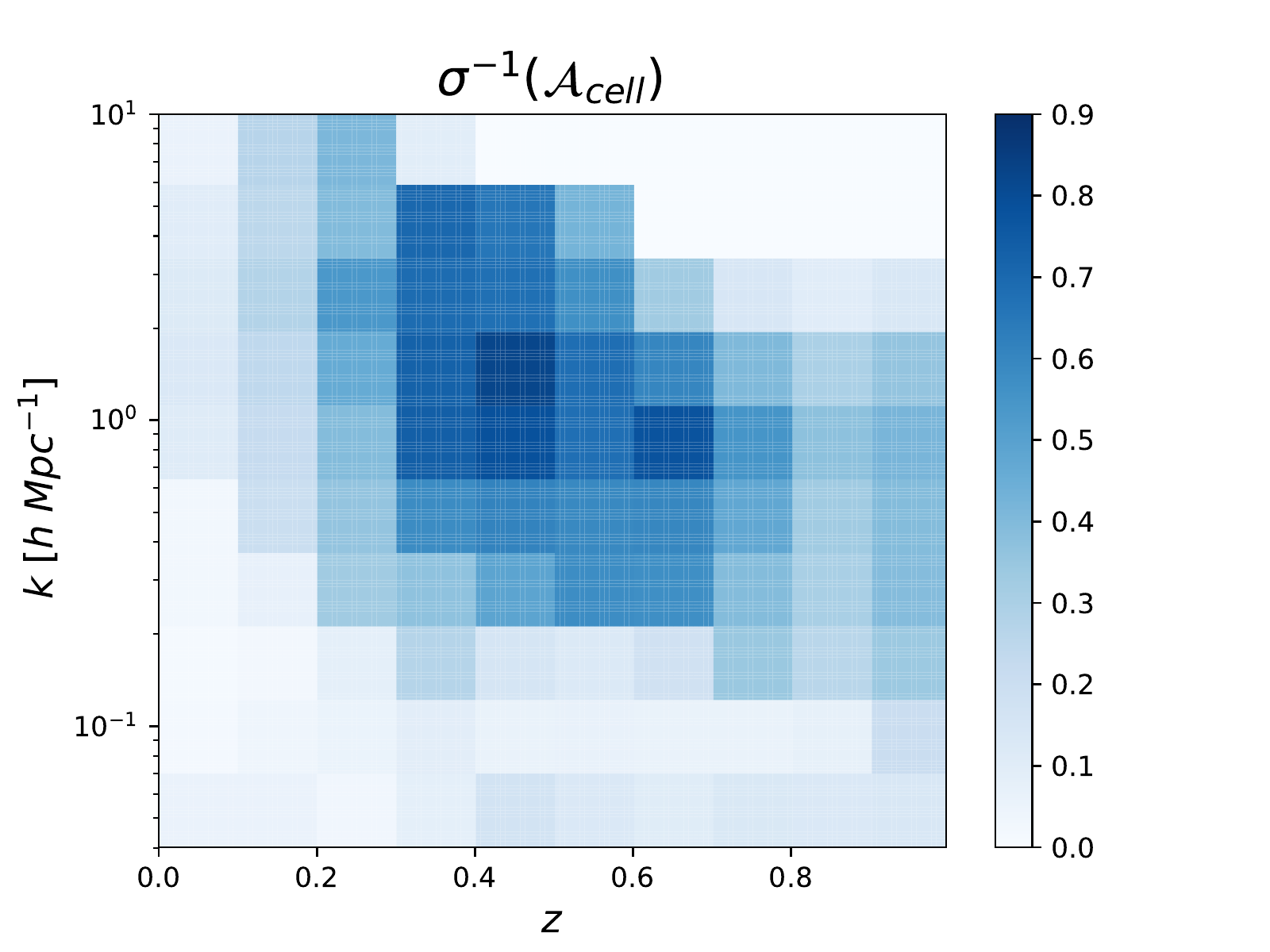}
        \includegraphics[width=5.5cm]{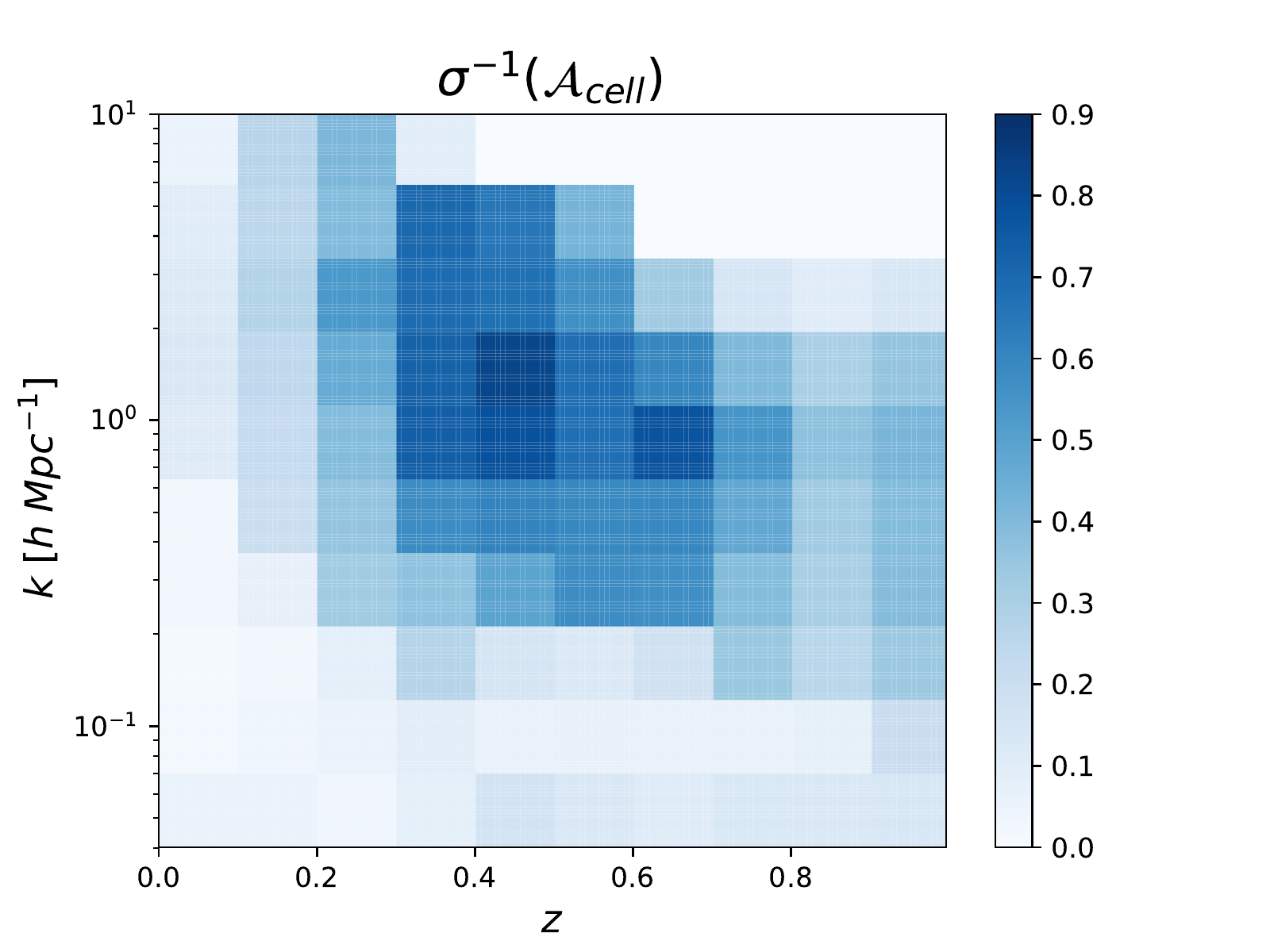} \\
        {\bf k-cut (redshift-independent)}\\
        \includegraphics[width=5.5cm]{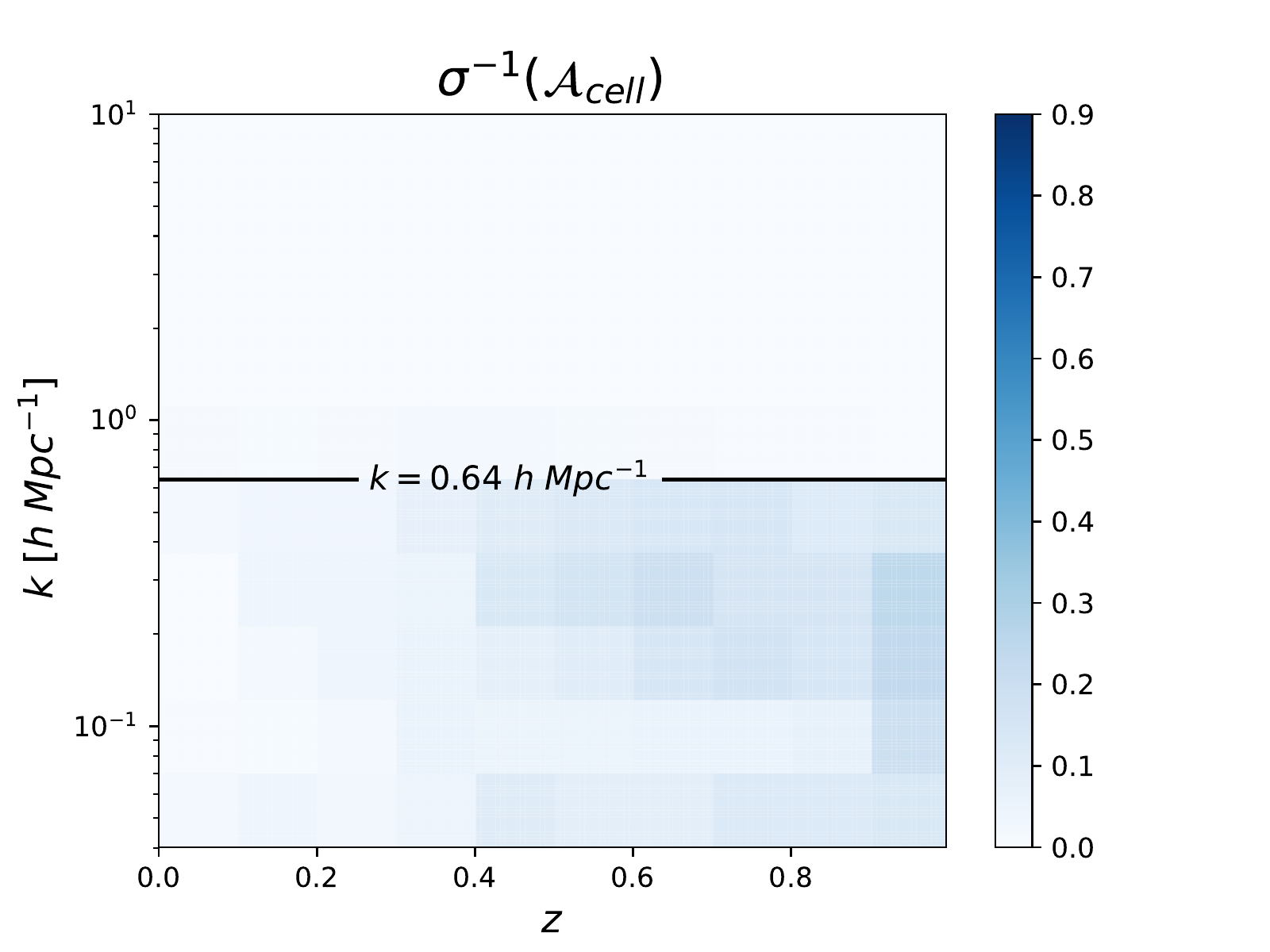}
        \includegraphics[width=5.5cm]{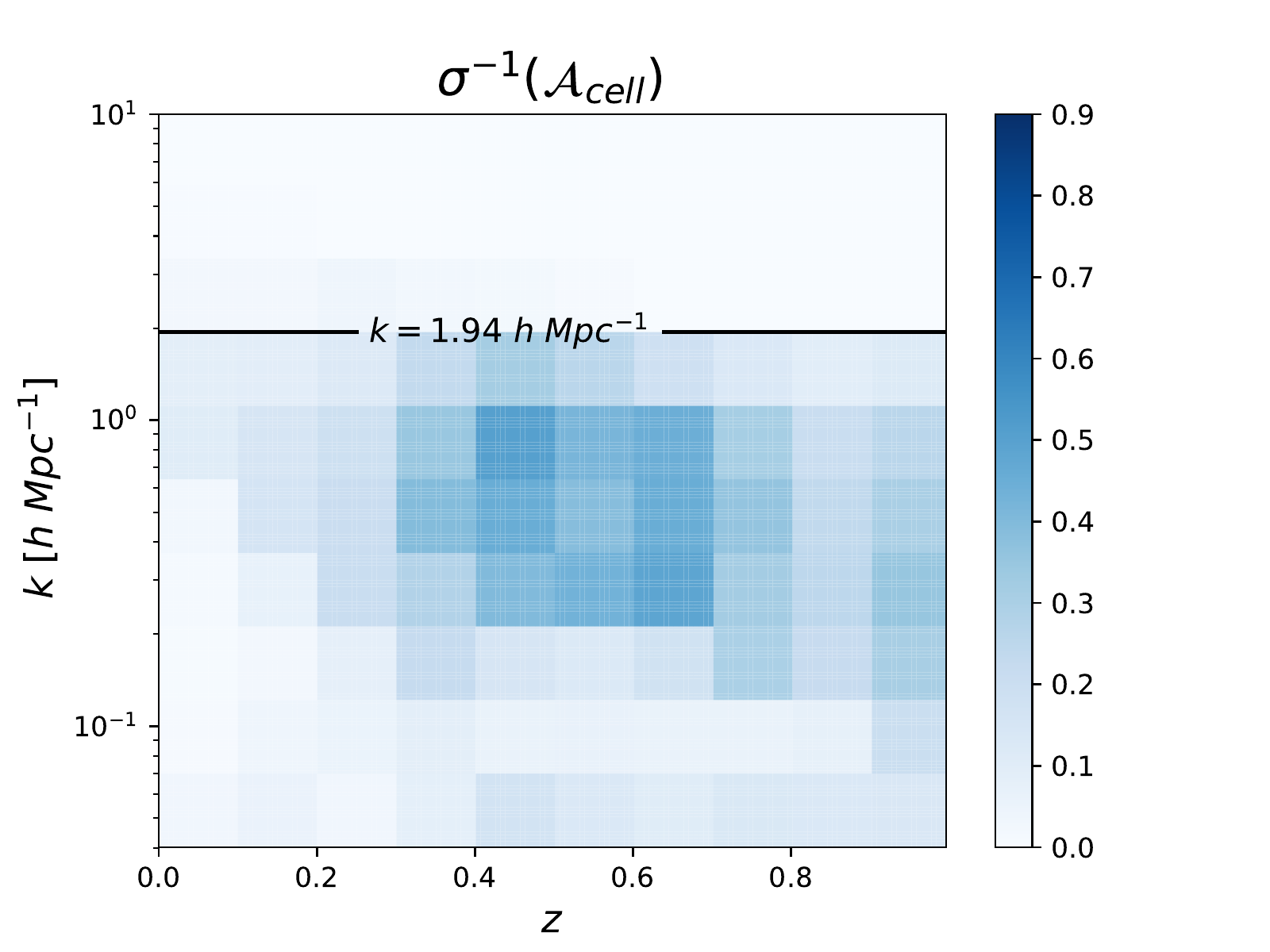}
        \includegraphics[width=5.5cm]{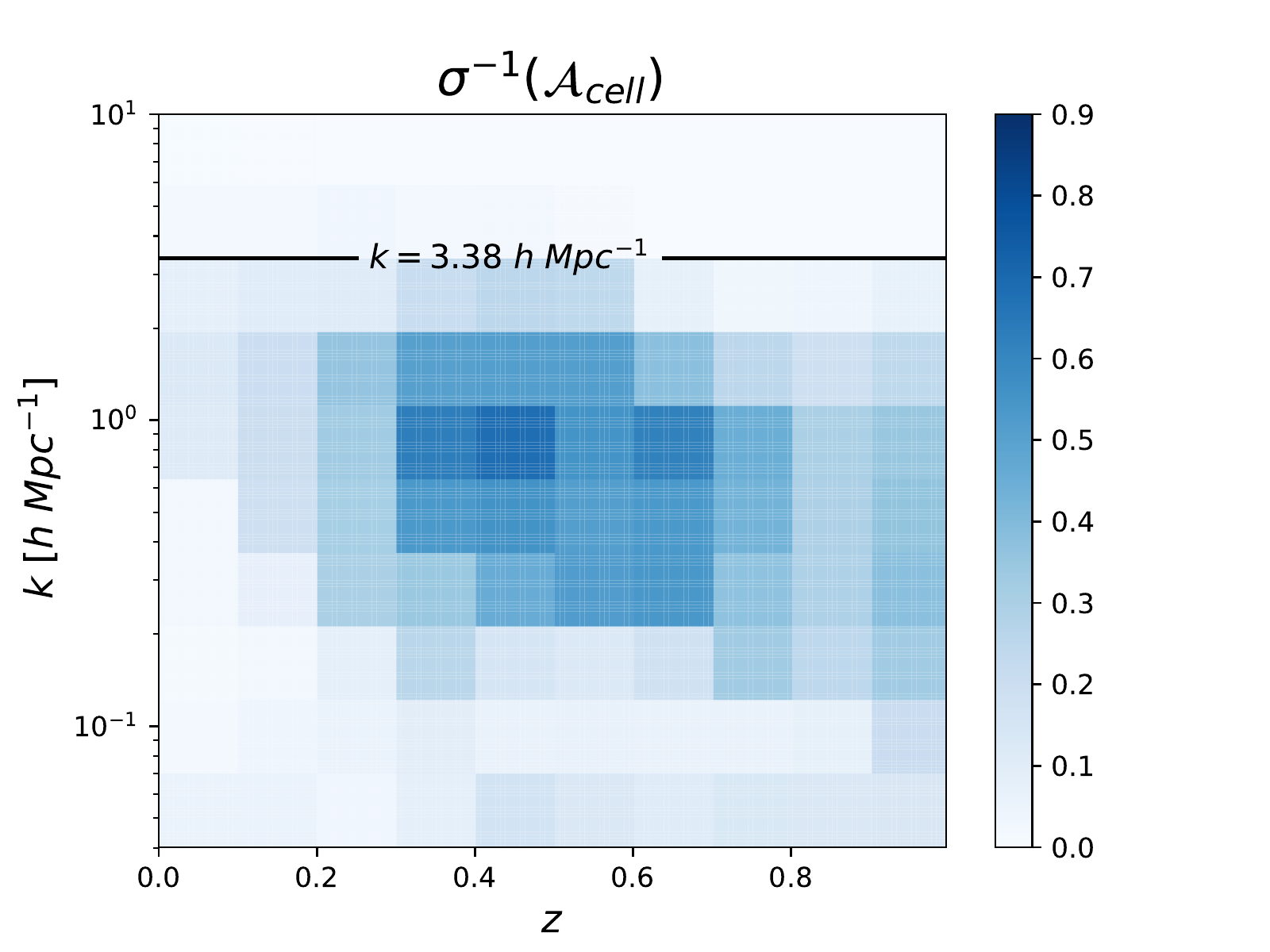}\\
        {\bf k-cut (redshift-dependent)}\\
		\includegraphics[width=5.5cm]{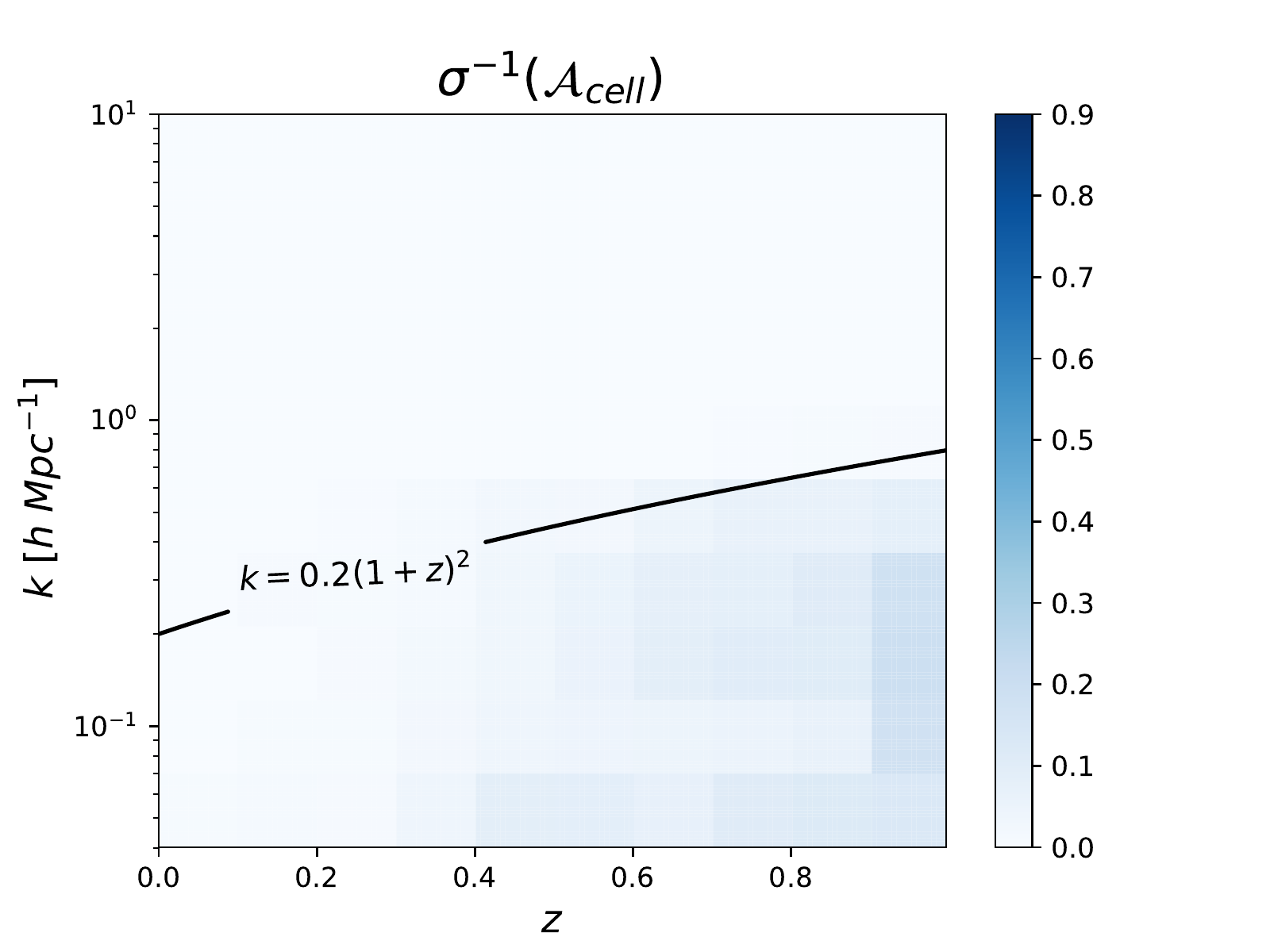}
        \includegraphics[width=5.5cm]{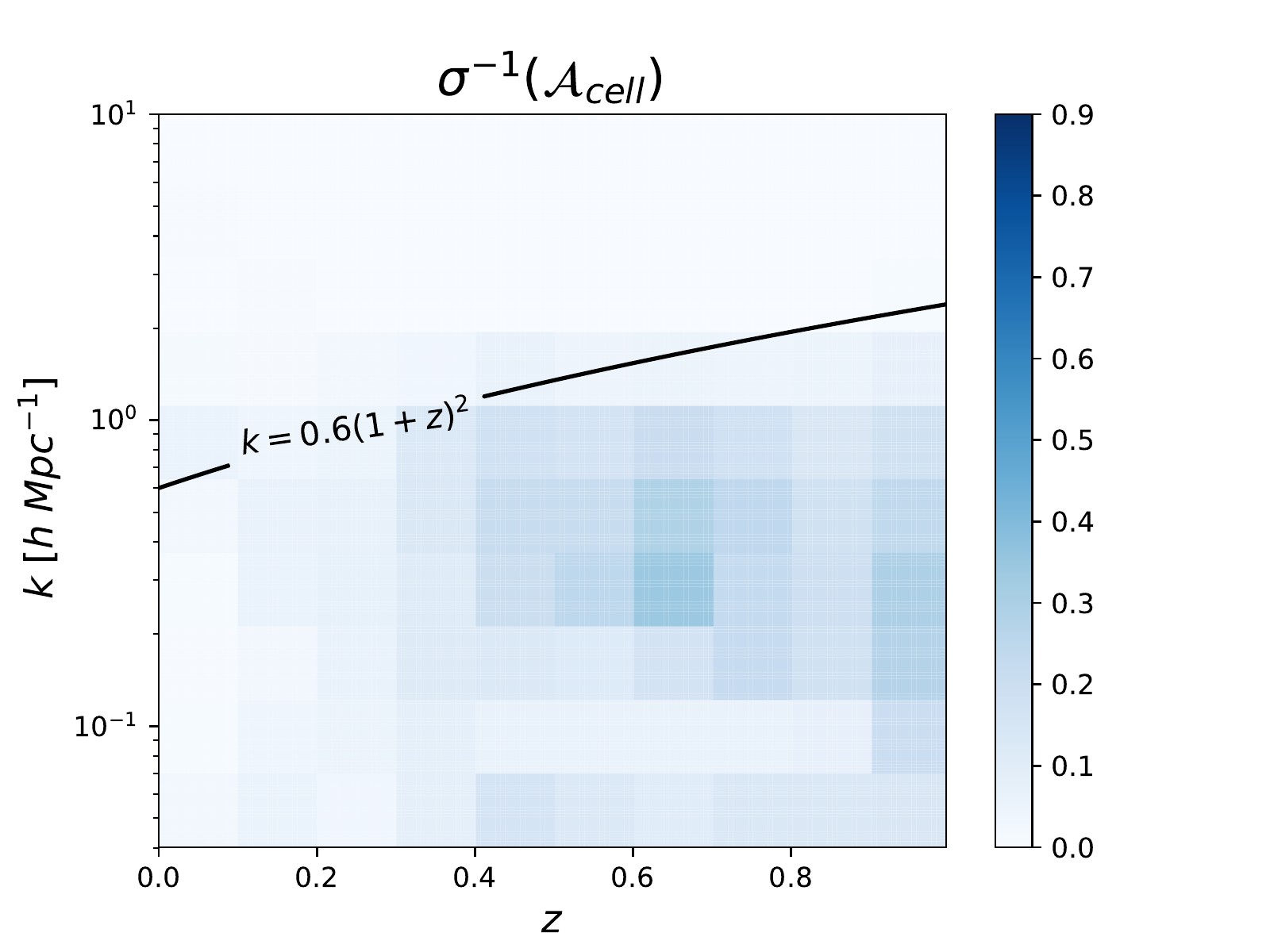}
        \includegraphics[width=5.5cm]{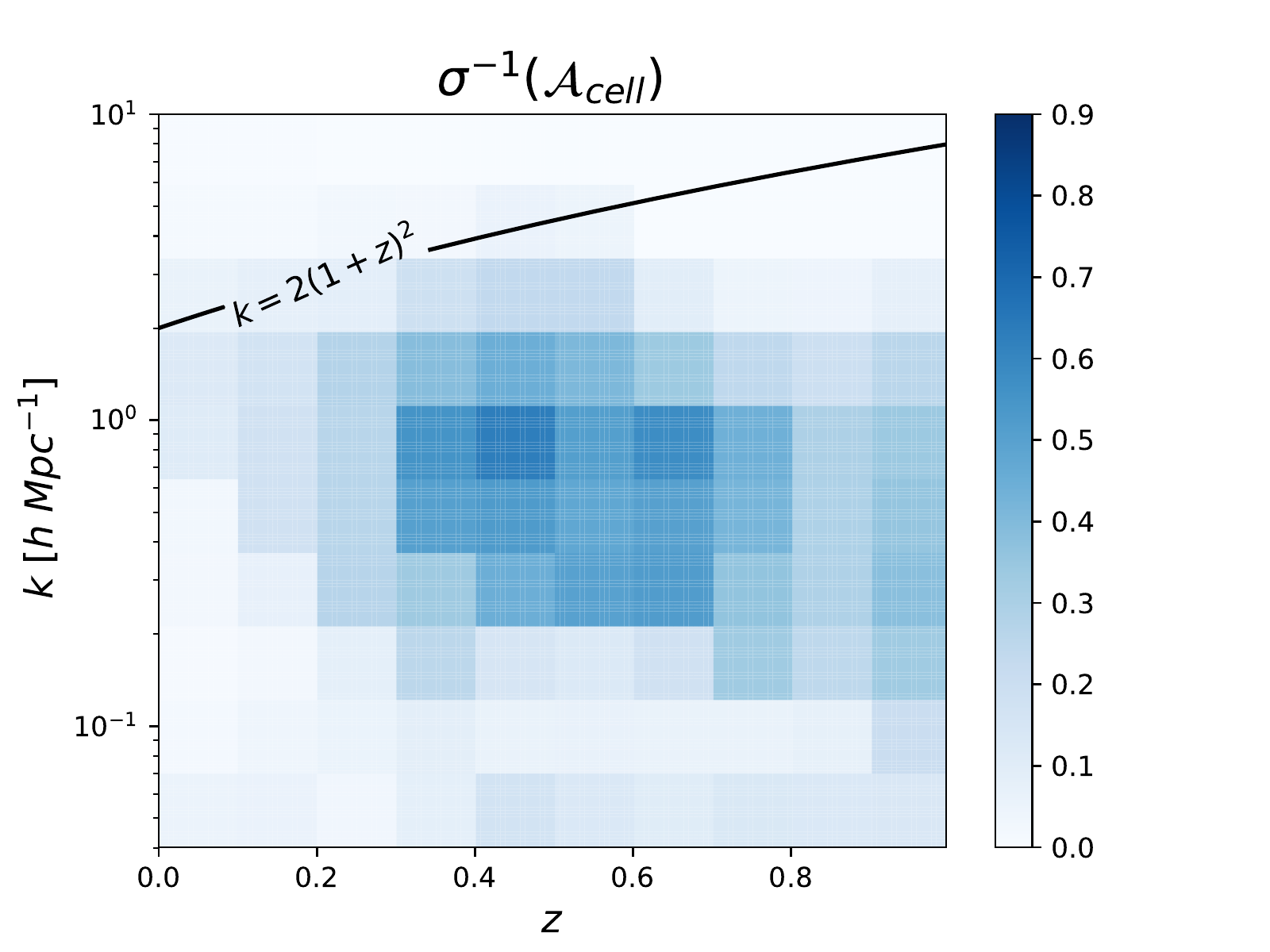}
		\caption{The Fisher matrix predictions for the inverse error on the measured amplitude of each power spectrum cell, $\sigma ^{-1} \left( \mathcal{A} \right)$, using different techniques. A given technique is sensitive to regions where $\sigma ^{-1} \left( \mathcal{A} \right)$ is high. {\bf Top left:} standard $C_\ell$ approach. {\bf Top right:} BNT weighting with no $\ell$-cut. BNT reweighing alone should not change the total sensitivity and there is at most a $0.02 \%$ fractional in any cell relative to the standard approach due to imprecisions in our numerical implementation {\bf Center row:} $k$-cut lensing with target $k_{cut}$ of the form $k_{cut}= A_{cut}$. {\bf Bottom row:} $k$-cut lensing with target $k_{cut}$ of the form $k_{cut}= A_{cut} (1+z) ^2$. $k$-cut cosmic shear efficiently removes sensitivity to the power spectrum above the desired $k$.}
		\label{fig:f1}
	\end{minipage}
\end{figure*}

 Figure~\ref{fig:f1} shows the inverse error on the amplitude of power spectrum cells for these six different cosmic shear analyses. Cosmic shear is most sensitive to dark blue regions. 
\par With the standard $C_\ell$ approach, shown in the top left, $\sim 50 \%$ of the signal comes from hard to model scales above $k_{\text{cut}} = 1$ $h$ $\text{Mpc} ^ {-1}$. The top right panel shows the case where we have applied $BNT$ re-weighting with no angular scale cuts. As expected, this had no effect on the sensitivity compared to the standard case. 
\par Finally in the last two rows we plot the sensitivity of $k$-cut cosmic shear with different target $k$-cuts. For all the cuts considered the sensitivity to regions above the target cut is dramatically reduced to essentially zero sensitivity. This is true even when photometric redshift errors are included, as is the case in our analysis. 
\par
\begin{figure}
   \centering
    \vspace{2mm}
    \includegraphics[width=90mm]{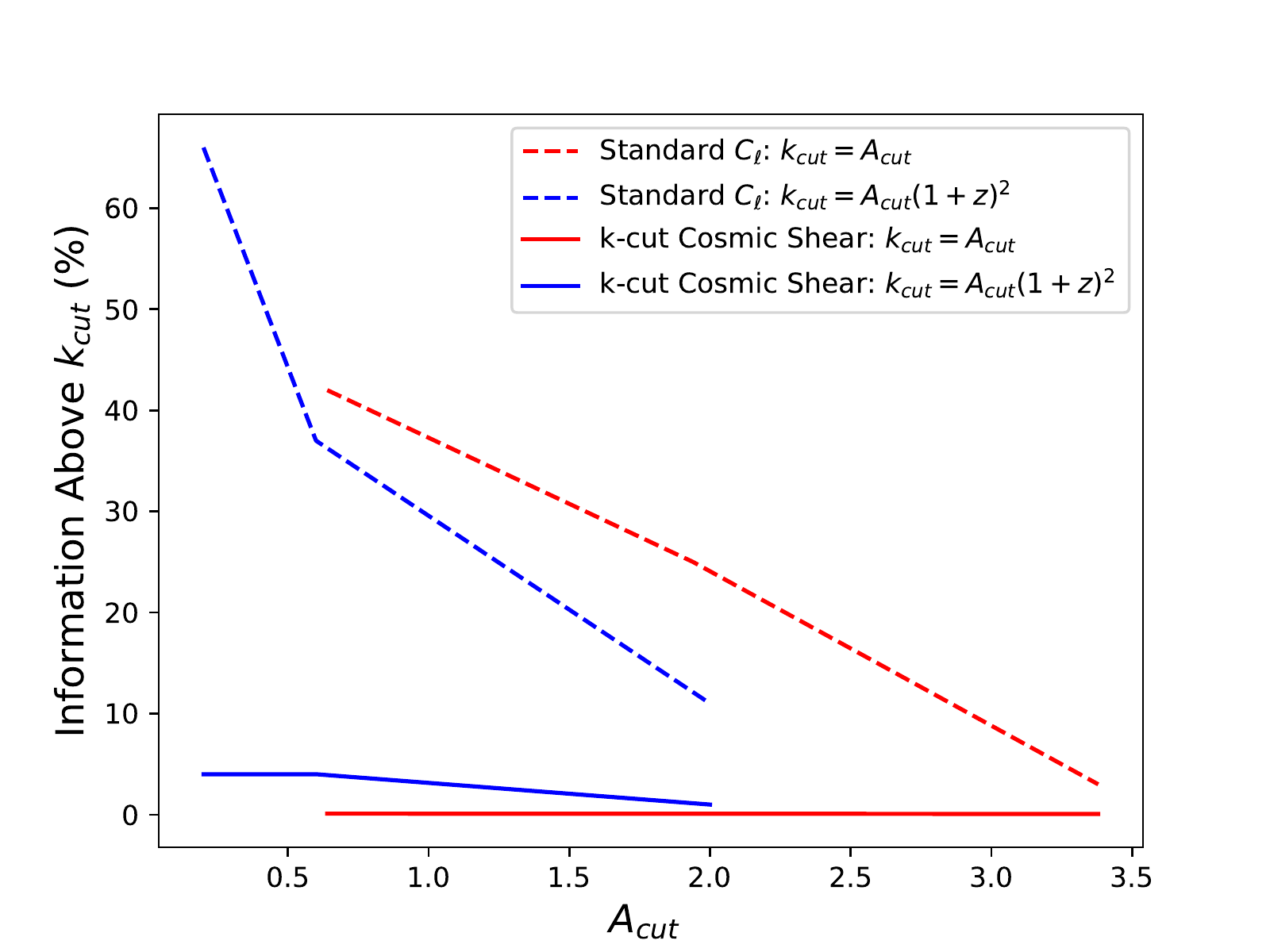}
    \caption{Fraction of the information coming from above the desired cut scale using the standard $C_\ell$ approach and $k$-cut cosmic shear. The information fraction is defined as the sum of the inverse errors (see equation~\ref{eqn:error}) on the cells above the cut relative to sum over all cells. $k$-cut cosmic shear removes nearly all sensitivity to small scales, while in the standard analysis a significant fraction of the signal comes from above the cuts.}
    \label{fig:info}
    \end{figure}

 The reduction in sensitivity to small scales is summarized in Figure~\ref{fig:info}. We plot the fraction of the matter power spectrum information that comes from scales above the cut. This is defined as the sum of the inverse errors (see equation~\ref{eqn:error}) on the cells above the cut relative to sum over all cells.
\par For all the cuts we considered, never more than $5\%$ of the information comes from scales above the target cut with $k$-cut cosmic shear. This is in contrast to the standard $C_{\ell}$ approach where up to $60 \%$ of the structure information comes from scales above the target cut.
\par
\begin{figure}
   \centering
    \vspace{2mm}
    \includegraphics[width=90mm]{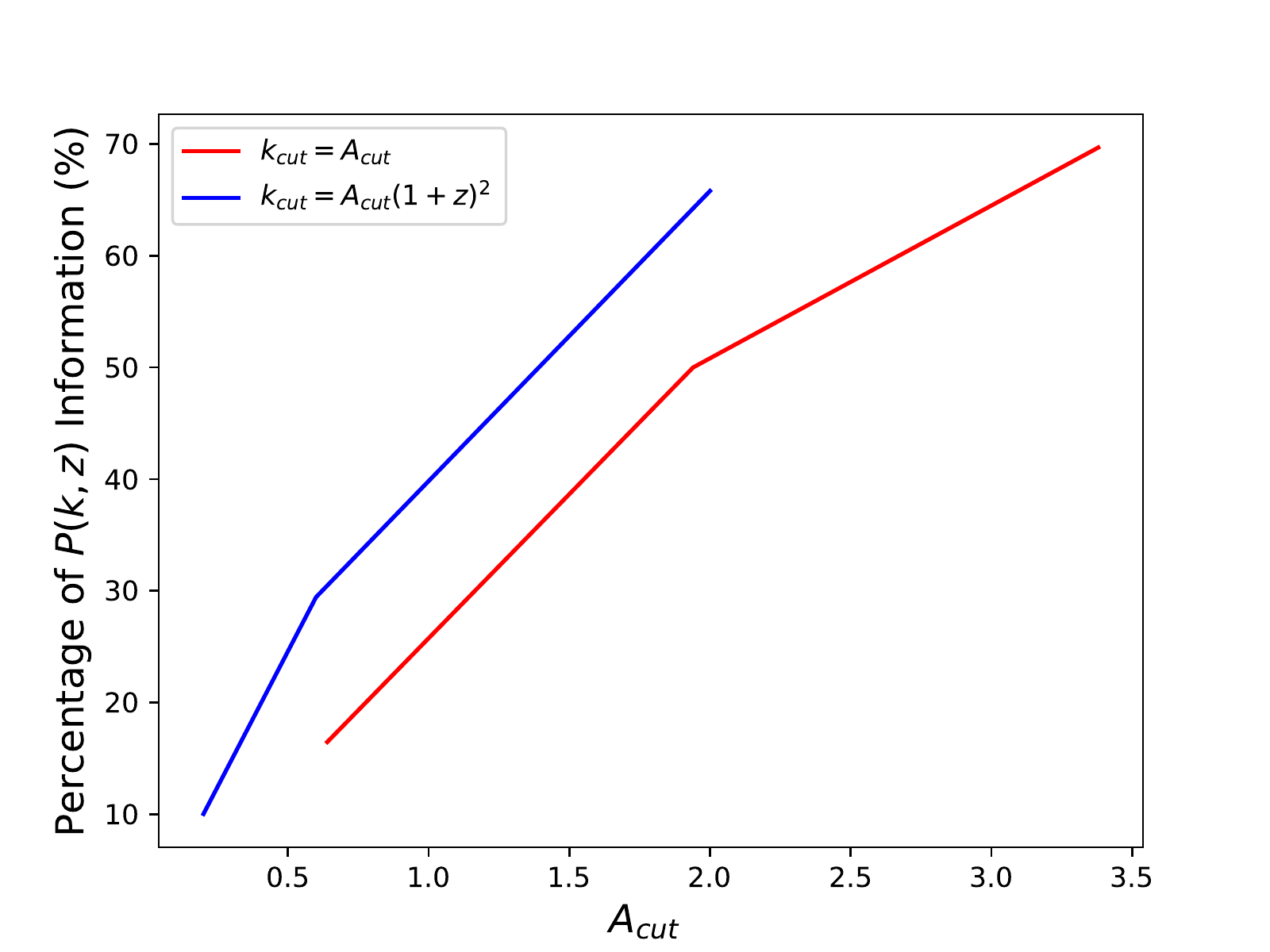}
    \caption{Fraction of the power spectrum information captured by $k$-cut cosmic shear relative to the standard $C_\ell$ approach. The information is defined as the sum of the inverse errors (see equation~\ref{eqn:error}) on the cells shown in Figure~\ref{fig:f1}. Although a large share of the power spectrum information is lost using $k_{cut}$ cosmic shear, by comparing with Figure~\ref{fig:w0}, we see that most of the information about the dark energy equation of state, $w_0$, is retained.  For example, when we take $k_{\text{cut}} = 1.94$ $h$ $\text{Mpc} ^ {-1}$, the size of error on $w_0$ only increases by $50 \%$, even though $70 \%$ of the power spectrum information is lost.}
    \label{fig:p_k_info}
    \end{figure}

\begin{figure}
   \centering
    \vspace{2mm}
    \includegraphics[width=90mm]{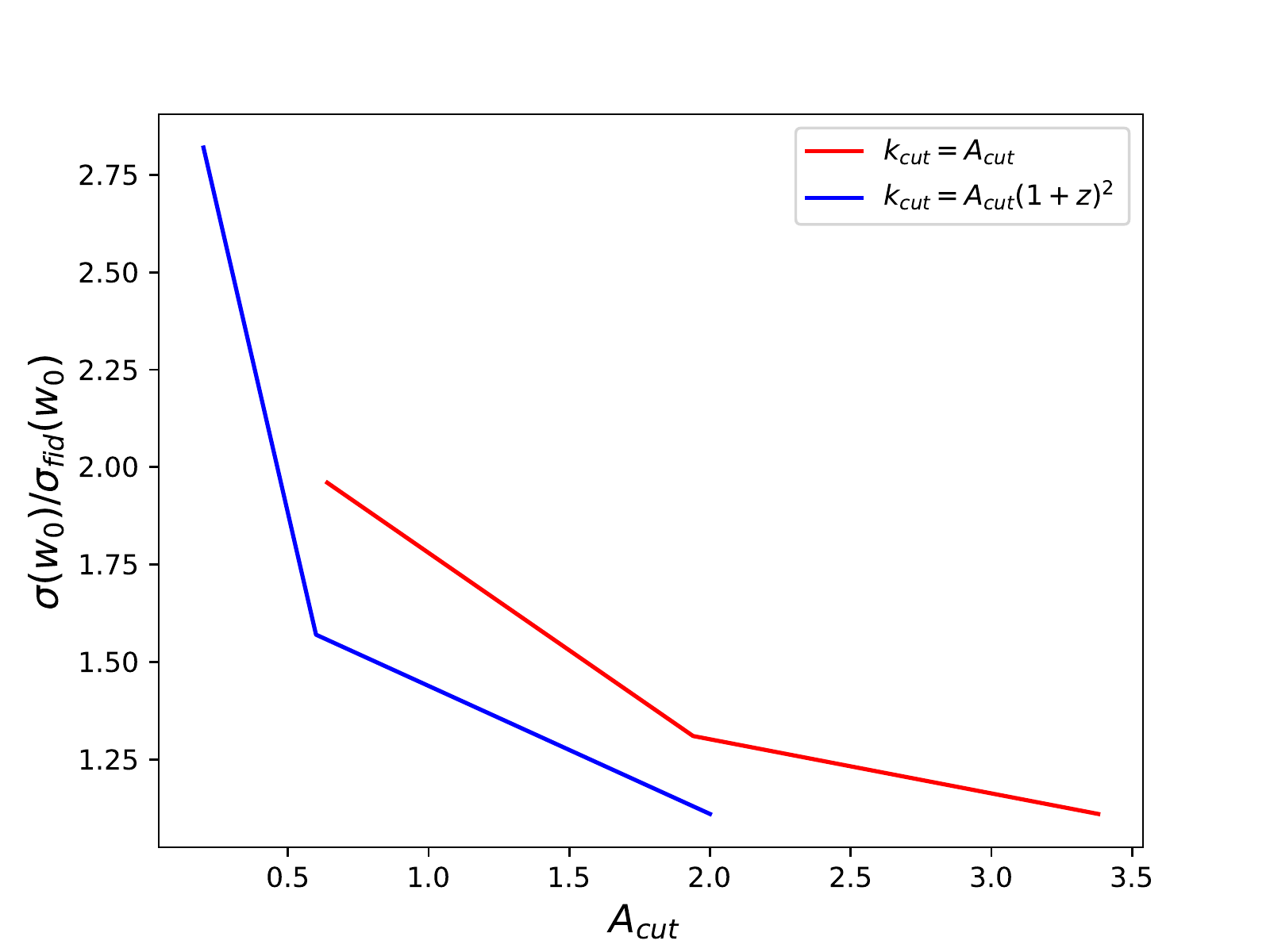}
    \caption{Size of $1 \sigma$ marginalised Fisher constraints on the dark energy equation of state, $w_0$, relative to the standard approach. In all but the most extreme case, where a very aggressive $k$-cut is used, a significant fraction of the sensitivity to small scales can be cut without degrading the $w_0$ constraint by more than a factor of $2$. As expected, applying the BNT transformation with no $k$-cut does not result in a loss of information.}
    \label{fig:w0}
    \end{figure}

 In Figure~\ref{fig:p_k_info} we plot the fraction of the power spectrum information retained using different $k$-cuts, relative to the standard approach. The information is defined as sum of the inverse errors on the power spectrum cells. When we take $k_{\text{cut}} = 1.94$ $h$ $\text{Mpc} ^ {-1}$, $70 \%$ of the power spectrum information is lost. For all cuts we considered, $>35 \%$ of the power spectrum information was lost. However, most of the constraining power on the dark energy equation of state is retained (see Figure~\ref{fig:w0}), because information about this parameter comes from large scales (small-$k$) in the power spectrum~\cite{copeland2018impact} and from the background geometry~\cite{taylor2018preparing}.
\par In Figure~\ref{fig:w0} we show the $k$-cut cosmic shear Fisher constraints on $w_0$, relative to the standard $C_\ell$ approach. Cutting scales does result in some loss of constraining power, but in all but the most extreme case that we considered, this never degrades the constraint on $w_0$ by more than a factor of 2. For examples cutting scales above $k_{\text{cut}} = 1.94$ $h$ $\text{Mpc} ^ {-1}$ results in a $31 \%$ increase on the size of the error. Meanwhile with the most aggressive cut that was considered -- where we removed sensitivity to all non-linear scales taking $k_{\rm cut} = 0.2(1+z) ^2$ -- the size of the error increases by a factor of $2.8$.

\section{Outlook and future prospects}
We have shown that $k$-cut cosmic shear is a clean and efficient way to remove sensitivity to small scales. Testing a modified gravity model with $k$-cut cosmic shear would require just three pieces of information from the theoreticians. These are:
\begin{itemize}
\item{The expansion history, that is the radial co-moving distance as a function of the redshift $r \left( z\right)$ which enters into the lensing kernel.}
\item{The matter power spectrum using any technique.}
\item{A breakdown scale $k \left(z \right)$ above which the power spectrum calculation is no longer sufficiently accurate. Determining the breakdown scale is survey-dependent and more work is needed in this area.}
\end{itemize}
Crucially this method would not require a non-linear model for small-scale matter power spectra, or a baryonic feedback model etc. above the cut-off scale (in contrast to standard cosmic shear).  
\par There are a few additional considerations which must be addressed before applying $k$-cut cosmic shear to data:
\begin{itemize}
\item{{\bf Intrinsic Alignments (IA)}: Since the IA signal is generated with a different kernel from the gravitational shear signal, $k$-cut cosmic shear does not remove small scales from the IA contribution. Nevertheless this should not be a major concern because the IA contribution is (i) primarily sensitive to large scales through tidal distortions induced by massive halos and (ii) already very sub-dominant to the shear signal. Making this precise is left to a future work.}
\item{ {\bf Covariance Matrix:} Testing each theory of gravity may require cosmology dependent covariances~\cite{eifler2009dependence}. Since $k$-cut cosmic shear is insensitive to small non-Gaussian scales, it would be interesting to re-examine whether analytic Gaussian-covariances are sufficient when removing small scales, or if cheap log-normal simulations~\cite{xavier2016improving,troxel2017dark} are sufficient. This may be a further advantage of $k$-cut cosmic shear. Sidestepping the issue altogether with likelihood-free methods also looks like a promising technique~\cite{alsing2018massive}.}
\item{ {\bf Mode Coupling:}} In linear theory each $k$-mode evolves independently, but nonlinear and baryonic corrections couple $k$-modes smearing modeling errors across a wide range in $k$. This is why the accuracy of leading emulators and halo model codes only vary slowly across a large range in $k$. For example the stated accuracy of {\tt HALOFIT} \cite{halofit} is $5 \%$ for $k\leq 1 h \text{ Mpc} ^{-1}$ and $10 \%$ for $k\leq 10 h \text{ Mpc} ^{-1}$. Meanwhile {\tt COSMIC EMU} \cite{heitmann2013coyote} report $4 \%$ accuracy for $k \in  [0.1 h \text{ Mpc} ^{-1}, 10 h \text{ Mpc} ^{-1}]$ and {\tt HMCode} \cite{mead2015accurate} report $5 \%$ accuracy for $k \in  [0.1 h \text{ Mpc} ^{-1}, 10 h \text{ Mpc} ^{-1}]$. Nevertheless it is generally the case that small $k$-modes are modeled less accurately than large-$k$ and it should still be possible to define a suitable cut scale. However this issue is also a worry for standard cosmic shear analyses. 
\end{itemize}
\par Addressing these remaining issues should be a priority since $k$-cut cosmic shear provides a way to enable a test of gravity, free from issues of uncertain small-scale bias. 

\section{Acknowledgements}
The authors would like to thank the developers of all public code used in this work. We are also grateful for constructive conversations with Eric Huff and Jason D. McEwen. PLT is supported by the UK Science and Technology Facilities Council. TDK is supported by a Royal Society University Research Fellowship. The authors acknowledge the support of the Leverhulme trust.

\section{Appendix}
The shear spectrum, $C_{\ell}^{\gamma \gamma}$, is given by: 
\begin{equation} \label{eq:c_l}
C_{\ell}^{\gamma \gamma} \left( \eta_i, \eta_j \right) =  \frac{9 \Omega_m ^ 2 H_0 ^ 4}{16 \pi^4 c^ 4 }\frac{\left( \ell + 2 \right)!}{\left( \ell - 2 \right)!} \int \frac{\text{d} k}{ {k} ^2} G_{\ell}^\gamma \left( \eta_1, k \right) G_{\ell}^\gamma \left(\eta_2, k \right) ,
\end{equation}
where $\Omega_m$ is the fractional energy density of matter, $c$ is the speed of light in vacuum and $H_0$ is the value of the Hubble constant today. $\eta_{i,j}$ label tomographic bins $i$ and $j$. The $G$-matrix is:
\begin{equation} 
\label{eq:G}
\begin{aligned}	
G_{\ell} ^ \gamma \left( \eta_i , k \right) \equiv \int \text{d}z_p \text{d} z' \text{ }  &n \left(z_p \right) p \left(z' | z_p \right) \\ & \times W_i U_{\ell} \left(r \left[ z' \right], k \right)  
\end{aligned}
\end{equation}
where $r[z]$ is the co-moving distance at a redshift $z$. The weight function, $W_i$, is a top hat function over redshift bin $I$. We assume $10$ redshift bins with an equal number of galaxies in each bin. The radial distribution of galaxies denoted by $n(z)$ is taken as: 
\begin{equation}
n \left( z \right)  = \left( z/ z_{e} \right) ^2 e^ {- \left( z / z_e \right) ^ {3/2}} ,
\end{equation}
with $z_e = 0.9 / \sqrt{2}$. 
The Gaussian photometric smoothing term, $p \left( z|z' \right)$, is:
\begin{equation} \label{eq:photo error}
p \left( z | z_p \right) \equiv \frac{1}{2 \pi \sigma_z \left(z_p \right)} e ^{- \frac{ \left( z -c_{cal} z_p + z_{bias} \right) ^2  } {2 \sigma_{z_p}} },
\end{equation}
with $c_{cal} = 1$, $z_{bias} = 0$ and $\sigma_{z_p} = A \left(  1 + z_p \right)$ with $ A = 0.05$ \cite{ilbert2006accurate}.
Meanwhile the $U$-matrix is:
\begin{equation} \label{eq:U}
U_{\ell} \left(r[z], k \right) \equiv \int ^ r _0 \text{d} r' \text{ } \frac{ \left(r - r' \right)}{a \left(r' \right) r r'} j_{\ell} \left( k r' \right) P^{1/2 }\left(k ; r' \right),
\end{equation}
where $a$ is the scale factor, $j_\ell(kr)$ are the spherical Bessel functions and $P(k;r)$ is the power spectrum. We use {\tt CAMB}~\cite{camb} to generate the linear power spectrum, {\tt Halofit}~\cite{halofit} to generate the nonlinear part. We assume a fiducial cosmology of $\left( \Omega_m, \Omega_k, w_0, \Omega_b,  h_0, n_s, A_s,  \tau  \right) = \left(0.32,\text{ }  0.0,\text{ }, -1.0, 0.04,\text{ }  0.67,\text{ } 0.96,\text{ } 2.1 \times 10 ^9,\text{ } 0.08 \right)$ throughout.
We assume the Limber approximation for $\ell > 100$ in which case the $U$-matrix becomes:
\begin{equation} \label{eq:limber}
U_{\ell} \left(r, k \right) = \frac{ r - \nu \left( k \right) }{k a \left( \nu\left( k \right)  \right) r \nu \left( k \right)} \sqrt {\frac{\pi}{2 \left( {\ell} + 1/2 \right)}}  P ^ {1/2} \left( k, \nu\left( k \right)  \right),
\end{equation}
where $\nu\left( k \right)  \equiv \frac{{\ell}+ 1/2}{k}$. Throughout we take $\ell_{\rm max} = 5000$. The contribution to the spectrum caused by the random ellipticity of galaxies, called the shot noise, is given by:
\begin{equation} \label{eq:Noise}
N_\ell^{ e e} \left( \eta_1, \eta_2 \right) = \frac{\sigma_e ^2}{2 \pi ^ 2 \Delta \Omega n_{\text{eff}}}, 
\end{equation}
where $\sigma_e ^2$ is the variance of the intrinsic (unlensed) ellipticities of the observed galaxies. We use $\sigma_e  = 0.3$ throughout \cite{brown2003shear}. In our analysis we assume the survey area, $\Delta \Omega$, is 15,000 square degrees and we use an effective number density of galaxies, $n_{\text{eff}}$, of 30 galaxies per $\text{arcmin} ^2$.

\bibliographystyle{apsrev4-1.bst}
\bibliography{bibtex.bib}


\end{document}